\documentclass[aps,prl,twocolumn,floats,subeqn,showpacs,floatfix,a4paper]{revtex4}
\usepackage{epsfig}
\usepackage{color}
\usepackage{bm}
\usepackage{latexsym}
\usepackage{subeqn}

\begin{document}

\bibliographystyle{unsrt}

\newcommand{\half}{\mbox{\small $\frac{1}{2}$}}
\newcommand{\bra}[1]{\left\langle #1\right|}
\newcommand{\ket}[1]{\left|#1\right\rangle }
\newcommand{\acos}[1]{\mbox{a}\cos }

\title{Thermalization of Strongly Disordered Nonlinear Chains}
\author{Tsampikos Kottos$^{1,2}$ and Boris Shapiro$^3$}

\affiliation{$^1$Department of Physics, Wesleyan University, Middletown, Connecticut 06459, USA \\
$^2$MPI for Dynamics and Self-Organization, Bunsenstr.\ 10, 37073 G\"ottingen, Germany\\
$^3$Department of Physics, Technion - Israel Institute of Technology, Haifa 32000, Israel}

\begin{abstract}
Thermalization  of systems described by the discrete non-linear Schr\"odinger equation, in the strong disorder
limit, is investigated both theoretically and numerically. We show that introducing correlations in the disorder
potential, while keeping the ``effective'' disorder fixed (as measured by the localization properties of wavepacket
dynamics), strongly facilitate the thermalization process and lead to a standard grand canonical distribution of
the probability norms associated to each site.
\end{abstract}

\pacs{05.45.-a, 72.15.Rn, 63.20.Pw, 63.50.-x}

\maketitle


{\it Introduction --} The discrete nonlinear Schr\"odinger equation (DNLSE) continues to be an active area of research. 
It became a paradigm in the field of nonlinear dynamical systems \cite{Kevbook} and it occupies an important place in 
nonlinear optics \cite{LSCASS08} and the theory of ultra-cold atomic Bose-Einstein condensates \cite{PSbook}. Much of 
recent work is dealing with the analysis of a wavepacket spreading, i.e. only one site (or, perhaps, a compact group of 
a few) is excited and the time evolution of this initial excitation is followed. The focus of this activity is  the 
interplay of non-linearity and Anderson localization induced by a random potential \cite{review}. A different type of 
problem \cite{B10, RCKG00} is when, initially, all sites are excited, so that the amount of energy endowed to the system 
is proportional to its (infinite, in the thermodynamic limit) volume  \cite{note1}. Following the evolution of such initial preparation 
one can ask whether a large isolated system thermalizes. Specifically, one can consider a large but finite part of the
isolated system and ask whether it approaches a grand canonical distribution with some values of the temperature $T$ and 
chemical potential $\mu$.

The two types of problems, i.e. "spreading" vs "thermalization", are physically distinct, as emphasized in the work
of Basko \cite{B10}. In the former case, the system is excited locally (a microscopic excitation), whereas in the latter
the excitation is global i.e. macroscopic. In a finite size system the distinction between the two problems can become
blurred, if one does not specify how the system's energy scales with its size. It is possible, for instance, to
consider the case when the energy provided to the system remains fixed and finite, while the size of the system
increases \cite{MAPS09,S97} (the problem of a wave packet spreading belongs to this category). In such a case the
system, assuming it thermalizes, will reach a temperature which is inversely proportional to system's size and, thus,
approaches zero in the thermodynamic limit.

Here we consider genuine, "thermodynamic" thermalization, when the system's energy is proportional
to its size. We concentrate on the case of strong disorder, when in the absence of the non-linearity, all modes are 
strongly localized. We consider both uncorrelated and correlated random
potentials and show that introducing correlations in the disorder, while keeping the {\em effective} disorder
fixed (as it is measured by the spreading of an initially localized wavepacket in the absence of non-linearity),
strongly facilitate the thermalization process.

{\it Model--} The time-dependent DNLSE is given by
\begin{eqnarray}
i\frac{d\psi_n}{dt} = \epsilon_n \psi_n - v(\psi_{n+1} +
\psi_{n-1}) +\chi|\psi_n|^2\psi_n, \label{DNLS}
\end{eqnarray}
where $\epsilon_n$ is the $n$-th site energy, while the coupling constant $v$ is set below to unity. The nonlinearity 
strength $\chi$  and the time $t$ are measured, respectively, in units of $v$ and $v^{-1}$. The model (\ref{DNLS}) can 
be interpreted as a system of coupled nonlinear classical oscillators \cite{ELS85,B10}, with each oscillator being 
identified with the corresponding lattice site. We use ``site'' and ``oscillator'' interchangeably, and the same goes 
for ``site energy'' and ``oscillator frequency''. Below we will consider a finite lattice $1\leq n\leq N$ with periodic 
boundary conditions. To define the problem completely one has to specify the sequence of $\epsilon_n$'s as well as the 
initial preparation.

We will study three kinds of $\epsilon$-sequences: (A) The step-potential where the first $N/2$ sites are assigned on
-site energy $\epsilon_{\rm L}=0$, while the rest $N/2$ sites are having energy $\epsilon_{\rm R} =const \neq 0$; (B)
Uncorrelated random potential where the $\epsilon_n$'s are independent random variables chosen from a rectangular 
distribution of width $2W$, centered at zero; and (C) Correlated random potential where a random sequence of $\epsilon_n$'s, 
such as in (B), is created first and then it is convoluted according to: ${\tilde \epsilon_n} = {1\over \sqrt{D}} \sum_k 
\epsilon_k \left[ 0.5 (1+\cos[{2\pi(n-k)\over N}])\right]^{(N/l)^2}$ where  $D=\sum_k \left[(1+\cos(2\pi k/N))/
2\right]^{2(N/l)^2}$. This choice of $D$ ensures that the second moment of the correlated potential, $\langle 
\tilde{\epsilon}^2\rangle $, is equal to that of the corresponding uncorrelated one, $\langle \epsilon^2\rangle $. A
simple calculation shows that, for $N/l\gg 1$, the correlation function $\langle \tilde{\epsilon}_n 
\tilde{\epsilon}_{n+m}\rangle =\langle \epsilon^2\rangle \exp(-m^2/\zeta^2)$, where the correlation length is $\zeta= 
l\surd2/\pi$.

The initial wavefunction amplitudes $\{\psi_n\}$ associated to the $n-$th site, were chosen for all cases to be a
random number in the interval $\left[0.75,1.25\right]$, while the normalization ${\cal N} = \sum_n |\psi_n|^2=N$
was imposed. Note that this implies that, initially, the phases of all $\psi_n$'s are zero. In addition to the
conserved norm ${\cal N}$, the total energy of the system ${\cal H}$ is also conserved by the dynamics of Eq.~
(\ref{DNLS}):
\begin{eqnarray}
{\cal H} = \sum_n [\epsilon_n |\psi_n|^2 -v(\psi_n^* \psi_{n+1} + c.c) +\frac{1}{2}\chi |\psi_n|^4],
\end{eqnarray}
where the three terms on the r.h.s correspond, respectively, to potential, kinetic and interaction energy. The conserved
quantities were monitored in our simulations to insure accuracy of the $4$rth order Runge-Kutta scheme.

{\it Theory --} We will investigate the thermalization process for the system of oscillators described by
Eq. (\ref{DNLS}). We assume that the coupling between oscillators is weak. For the random cases, (B,C), this is ensured
by strong disorder, i.e. by the large parameter $W/v$. For the step potential there is no randomness and the effective 
coupling strength depends on the initial preparation. It turns out that the aforementioned preparation corresponds to a 
high temperature regime, in which case the effective coupling between oscillators is weak \cite{RCKG00,note1}. Under the weak 
coupling condition the thermalization criterion can be applied to an individual oscillator: in equilibrium, an oscillator 
with eigenfrequency $\epsilon_n$ should obey the grand canonical distribution (GCD) for its norm $I_n\equiv|\psi_n|^2$, 
with the phase of $\psi_n$ being completely random \cite{B10}:
\begin{eqnarray}
P_n(I_n)=\frac{1}{Z_n}\exp[\beta (\mu -\epsilon_n)I_n -\frac{1}{2}\beta \chi I_n^2].
\label{PI}
\end{eqnarray}
Above $\beta$ is the inverse temperature, while the partition function $Z_n$ is given by the integral
\begin{eqnarray}
Z_n&=&\int dI_n \exp[\beta(\mu -\epsilon_n)I_n -\frac{1}{2}\beta \chi I_n^2]\nonumber\\
   &=&\sqrt{\frac{\pi}{2\beta \chi}}\exp(\frac{\beta E_n^2}{2\chi})
      erfc (\sqrt {\frac{\beta}{2\chi}} E_n).
\label{ZI}
\end{eqnarray}
In Eq. (\ref{ZI}), we have introduced the notation $E_n\equiv\epsilon_n - \mu$.

Instead of reconstructing the full distribution, we will focus our investigation (except from the case (A)) at the
time average of the norm and its second moment of each individual oscillator:
\begin{equation}
\langle I_n(t)\rangle\equiv {1\over t}\int_0^{t} I_n(t') dt';\quad
\langle I_n^2(t)\rangle \equiv {1\over t}\int_0^{t} I_n^2(t') dt'
\label{I1I2}
\end{equation}
Their time dependence will allow us to conclude whether or not a specific oscillator has reached equilibrium.
In the long time limit, these quantities should be equal to their statistical average, with respect to the GCD
given by Eqs. (\ref{PI},\ref{ZI}). The latter averaging results in:
\begin{subequations}
\label{stataver}
\begin{eqnarray}
\langle I_n\rangle &=& \frac{1}{\beta \chi}(\frac{1}{Z_n}-\beta E_n),\\
\langle I_n^2 \rangle &=& \frac{1}{\beta \chi}(1 +\frac{\beta}{\chi} E_n^2
-\frac{1}{\chi Z_n} E_n).
\end{eqnarray}
\end{subequations}
For sufficiently weak nonlinearity, when $\chi\ll\beta E_n^2$, the partition function $Z_n$ can be expanded,
leading to:
\begin{eqnarray}
\langle I_n\rangle =\frac{1}{\beta E_n}(1 - 2\frac{\chi}{\beta E_n^2});\quad
\langle I_n^2 \rangle =\frac{2}{(\beta E_n)^2}(1- 5\frac{\chi}{\beta E_n^2}).
\label{expand}
\end{eqnarray}

\begin{figure}
\includegraphics[width=1\columnwidth,keepaspectratio,clip]{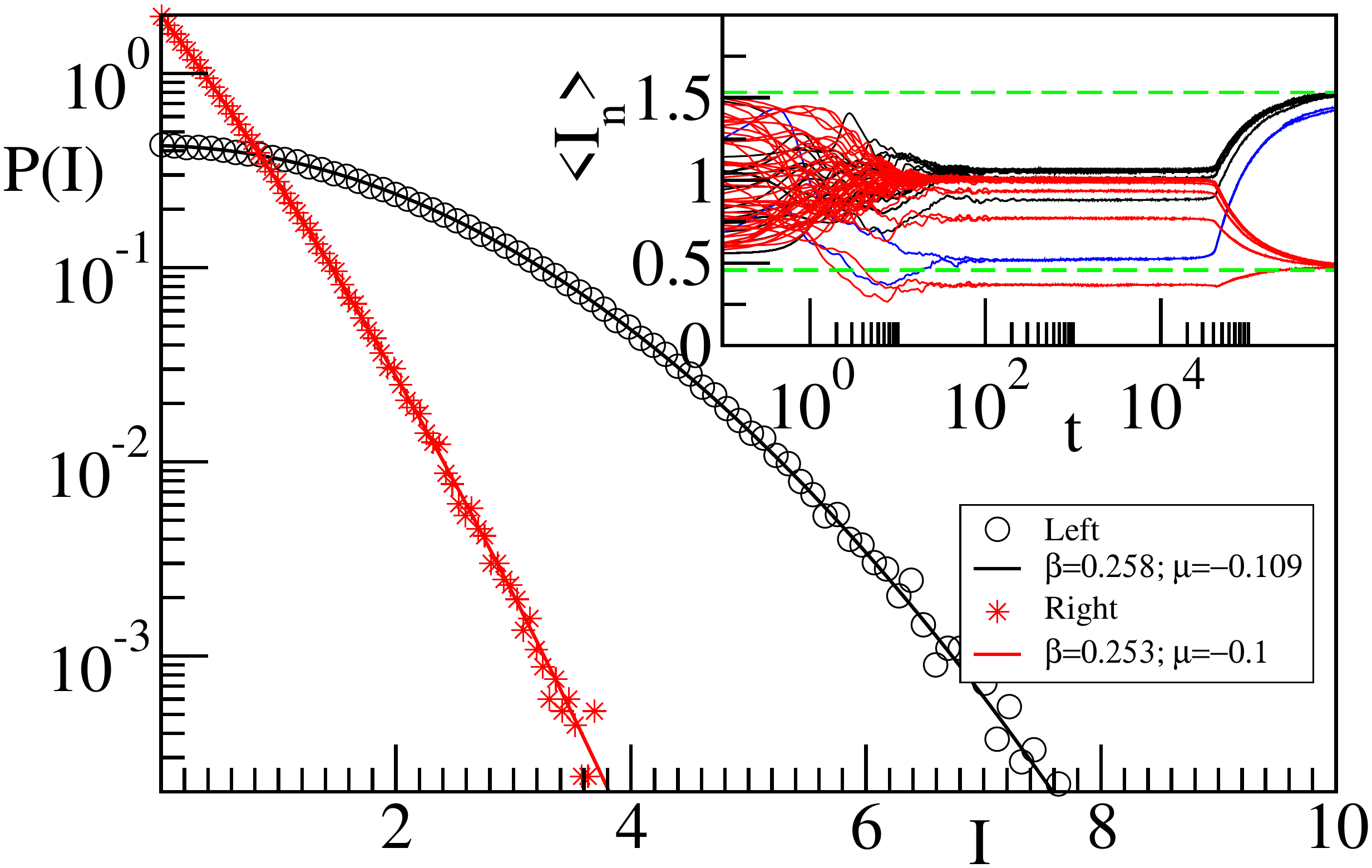}
\caption{(Color online) Distribution $P(I)$ of the site-norm $I$ for the case of a lattice with $N=100$ sites, $\chi=1$, and a 
step-potential with $\epsilon_L=0$ and $\epsilon_{\rm R}=7.5$. Black ($\circ$)/red ($\star$) colors correspond to left/right sites. 
The solid lines are best fits to Eq. (\ref{PI}). The fitted values ($\beta_L=0.258; \mu_L=-0.109$) for the left 
and ($\beta_R=0.253;\mu_R=-0.1$) for the right sub-chains respectively, agree nicely (within numerical accuracy
of the fit) with one-another. Inset: Evolution of the first moment of the time-averaged on-site norms $\langle 
I_n(t)\rangle$. The dashed lines are the theoretical predictions of Eq. (\ref{stataver}a) with $\beta=0.258$ and 
$\mu=-0.109$. Blue lines indicate border sites $n=1$ and $n=50$. } \label{fig1}
\end{figure}

{\it Numerical Analysis --} We now discuss our numerical results for each of the three on-site potential sequences. 
We start with the case (A) associated with the step potential. This is an instructive example of thermalization. We 
recall that the chain of $N$ sites is divided into two equal parts- the left and the right sub-chains with site 
energies $\epsilon_L=0$ and $\epsilon_R\neq 0$, respectively. The initial preparation with random on-site norms 
$I_n\sim 1$ evolve with time according to Eq (\ref{DNLS}). Thermalization of the system is achieved due to a significant
redistribution of the norm between various sites: ``flow of the norm'' from the right sub-chain (large site energies) 
towards the left sub-chain (zero site energies) occurs during the thermalization process. At equilibrium, a strong 
correlation between the norm at a specific site $n$ and the potential energy $\epsilon_n$ is developed, so that the 
requirement Eq. (\ref{stataver}) (or the more detailed condition for the full distribution) is satisfied. This scenario 
is confirmed by our numerical data for $\langle I_n\rangle (t)$, reported in the inset of Fig. \ref{fig1} for a chain 
of $N=100$ sites with $\chi=1$ and $\epsilon_{\rm right}=7.5$. Indeed, for short times (roughly up to $t=100$) the 
oscillators are exchanging norm among them and the various curves fluctuate wildly and cross each other. At larger
times they "get organized" into two groups of curves (with the exception of the few border sites between the two 
sub-chains). Note, however, that for a long interval of time (from $t\sim 100$ till $t\sim 5\times 10^4$) the two 
groups are very close to one another and do not change with time. For still longer times, they start diverging and 
eventually saturate at two different values for $\langle I_n\rangle $ which are predicted by Eq. (\ref{stataver}). 
We interpret this behavior as two-stage thermalization. First, oscillators within each sub-chain exchange norm and 
energy among them, with very little interaction between the two groups of oscillators. This rapid process (on time 
scales $t\sim 100$), leads to "internal" thermalization, associated to each sub-chain separately. Then, at much 
longer times, $t\sim 5\times 10^4$, the slow process of interaction between the two groups of oscillators becomes 
efficient and it eventually leads to a full equilibrium. Thus, the potential discontinuity, between the two sub-chains, 
serves as a ``bottleneck'' which slows down the total thermalization and provides the long-time scale for the
thermalization process.

We have also investigated the whole probability distribution $P(I_n)$ for the step potential chain. After an
initial transient $t\sim 10^5$, we have collected the norm $I_n$ of each oscillator at various times. In all 
cases more than $10^5$ data were used for statistical processing. We have found that the oscillators follow 
two distinct distributions, which are characterized by their on-site energy $\epsilon$ (i.e. left or right sub-
chain) according to the prediction of Eq. (\ref{PI}). We have observed that the oscillators that correspond to 
high on-site potential energy were first to thermalize while the ones on the left sub-chain, corresponding to 
$\epsilon_n=0$, achieve thermalization with a slower rate. A representative example of $P(I_n)$ is shown in the 
main panel of Fig. \ref{fig1}. For better statistical processing we have averaged the left and right distributions 
of all oscillators. We have found that both distributions $P_{\epsilon_L}(I)$ and $P_{\epsilon_R }(I)$ follow
nicely Eq. (\ref{PI}). The corresponding fitted left and right chemical potentials and temperatures are in 
excellent agreement with one another, indicating that the total chain reached global thermalization.

\begin{figure}
\includegraphics[width=1\columnwidth,keepaspectratio,clip]{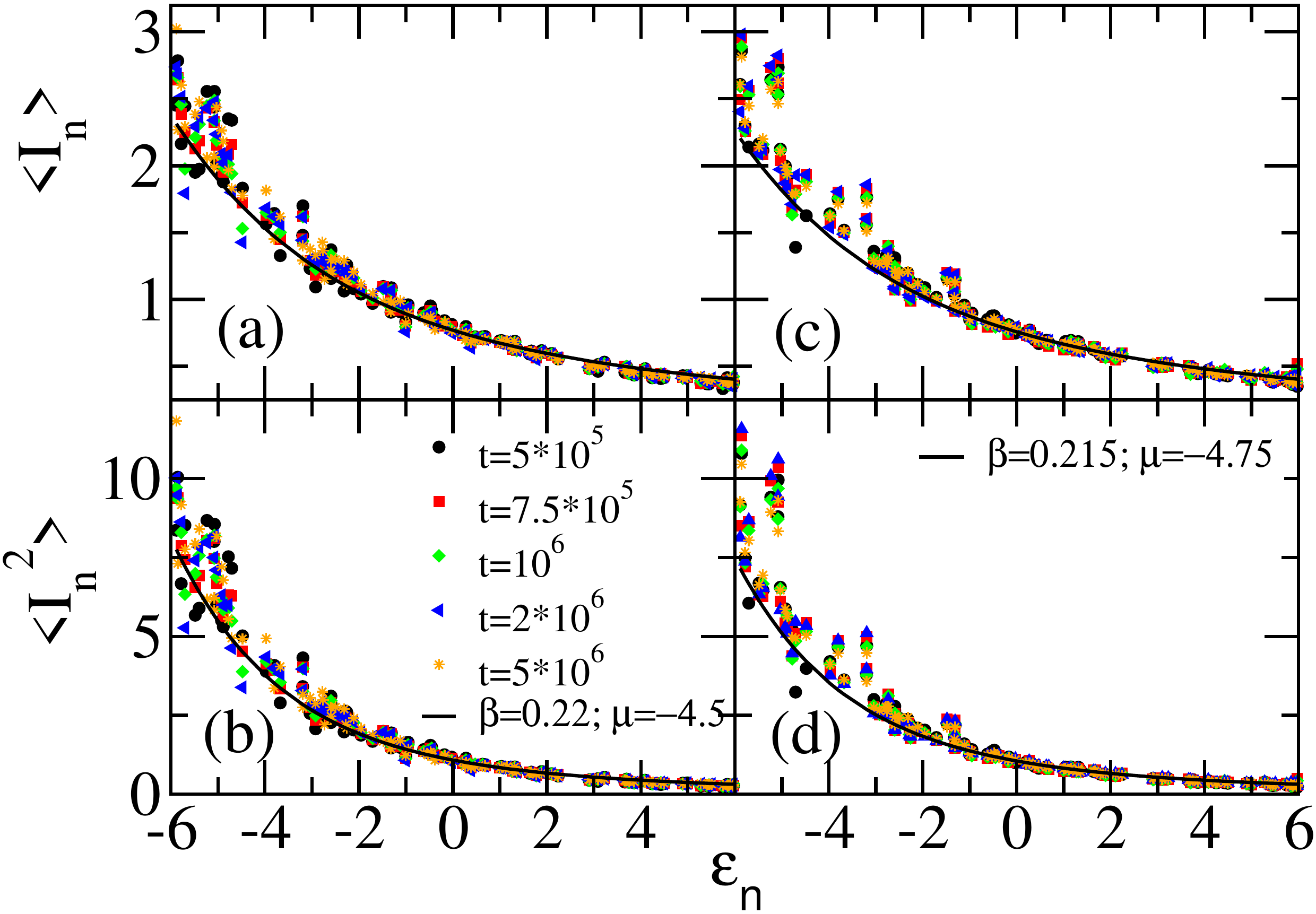}
\caption{(Color online) First $\langle I_n(t)\rangle$ (a,c) and second $\langle I_n^2(t)\rangle$ (b,d) moment of the on-site norm
vs. $\epsilon_n$ for an uncorrelated $\epsilon$-sequence with $W=6$. We considered a lattice with $N=100$ and
$\chi=1$. Subfigures (a,b) and (c,d) correspond to two different initial preparations. Solid lines correspond to
Eq. (\ref{stataver}a,b)
}
\label{fig2}
\end{figure}

(B) Next, we investigate the thermalization of a chain with uncorrelated on-site potential. We concentrate on 
the analysis of the first $\langle I_n\rangle$ and second moment $\langle I_n^2\rangle$ of the norm. We consider 
cases for which the coupling parameter, $\tau\equiv W^{-1}$ and the effective non-linearity strength $\rho\equiv 
\chi I_n /W $ are small \cite{B10}. We let the initial preparation to evolve for a long time, of the order of 
$5\times 10^5$ and then register the value of the norm $\langle I_n\rangle $, for all $N$ oscillators. Representative
results of $\langle I_n\rangle$ versus the site energies $\epsilon_n$'s for $W=6$, $N=100$, $\chi=1$ and two 
different initial preparations are shown in Fig. \ref{fig2}a,c for times $t=5\times 10^5; 7.5\times 10^5; 10^6; 
2\times 10^6; 5\times 10^6$. Such plot provides a check for the validity of Eq.~(\ref{stataver}) which should 
hold in equilibrium. In Fig. \ref{fig2}b,d  we report also the second moment, $\langle I_n^2\rangle $. One can 
see that for high-energy sites $-1\leq \epsilon_n\leq 6$ both quantities, already for $t=5\times 10^5$, follow 
a regular curve and stay there for longer times. This part of the data is nicely fitted to the expression Eq.
(\ref{stataver}), with appropriate $\beta$ and $\mu$ values. Moreover, we have checked that for $\epsilon > 4$ 
the norm $\langle I_n\rangle $ and, thus, the nonlinearity parameter become sufficiently small for the simplified 
relations Eq. (\ref{expand}) to hold. We have found, for instance, that for these highest energy tails, the 
relation $\langle I_n^2 \rangle = 2\langle I_n\rangle ^2$ is satisfied quite accurately. This is an indication 
for a negative-exponential distribution (Eq.(\ref{PI}) without the nonlinear term).

We now turn to the low energy part, $\epsilon \leq -3$, of the data in Fig. \ref{fig2}. These sites do not show 
a clear sign of thermalization. The points do not fall on a regular curve but rather form a ``blob''. Moreover, 
in the course of time the points are ``jumping around'' with no clear indication of approaching a regular curve. 
We conclude that, while the high energy sites thermalize already for relatively short times, the low energy ones
do not thermalize for much longer times (of course, they well might thermalize for even longer times). Let us note 
that for the lowest energy sites the nonlinearity parameter $\rho$ reaches a value of about $1/2$, so that the 
weak nonlinearity condition \cite{B10} is not really satisfied.

\begin{figure}
\includegraphics[width=1\columnwidth,keepaspectratio,clip]{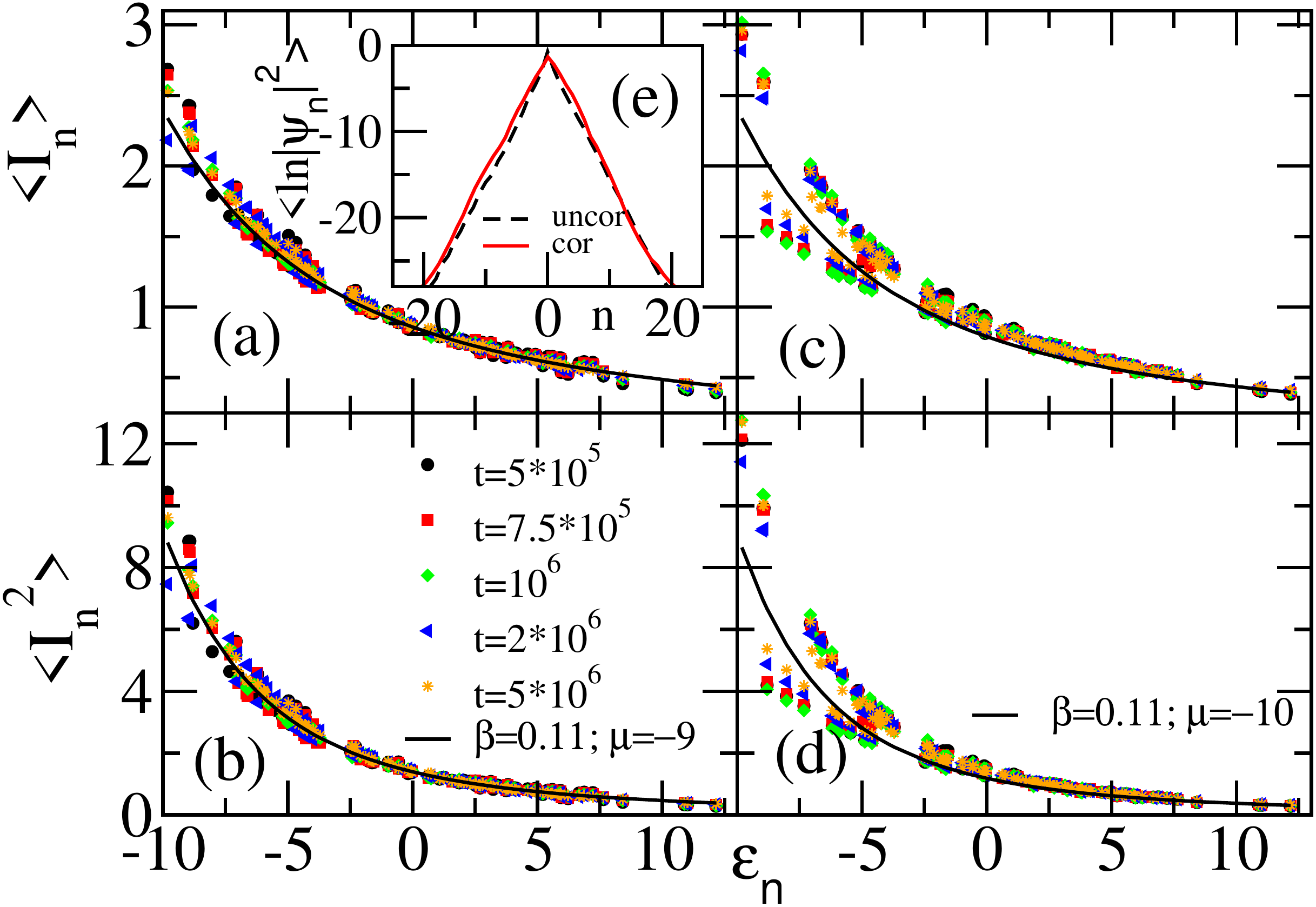}
\caption{(Color online) Same as in Fig. \ref{fig2}, but for a correlated ${\tilde \epsilon}_n$-sequence with $W=7$ and correlation 
length $\zeta=2.7$. Subfigures (a,b) and (c,d) correspond to two different initial preparations (the same as used 
for Fig. \ref{fig2}a,b and c,d respectively). In subfigure (e) we report the asymptotic wavefunction for a linear 
lattice with an uncorrelated (black dashed line) and a correlated (red solid line) $\epsilon$ -sequence with $W=6$ and $W=7$ (and $\zeta
=2.7$) respectively (the initial condition was a single site excitation).} 
\label{fig3}
\end{figure}

(C) Finally, we discuss the case of a correlated disorder and demonstrate that the thermalization process for 
this case is quite different from that in (B). A random correlated sequence of $\tilde{\epsilon}_n$ is created 
as explained above (from now on we remove the ``tilde''). The random correlated ensemble is characterized by 
the ``bare'' disorder $W$ and the correlation length $\zeta$. The first thing to note is that it would not be
appropriate to compare uncorrelated and correlated ensembles with the same $W$; the point is that introducing 
correlation while keeping $W$ fixed will, generally, change the "effective disorder" as measured, for instance, 
by spreading of an initially localized wave packet. Therefore the statement {\it correlations in the randomness 
facilitate thermalization} becomes meaningful only when the same {\it effective disorder} is kept for the cases 
(B) and (C). As a measure of effective disorder we chose the spreading of a particle (for $\chi=0$), initially 
placed at some site. More precisely, we consider the {\it ensemble averaged} $\langle \log|\psi_n|^2\rangle$, 
in the long time limit, where $n$ is the distance from the initial site. This quantity is plotted in Fig.
\ref{fig3}e for two different ensembles: uncorrelated disorder (case (B)) with $W=6$ and correlated disorder 
(case (C) with $W=7$ and $\zeta= 2.7$). Comparing the two curves one can conclude that, indeed, these two 
ensembles have the same effective disorder.

We are now in the position to study thermalization in a chain with correlated disorder and to meaningfully compare
the results with those in Fig. \ref{fig2} for the uncorrelated case. In Fig. \ref{fig3}a,b we report our numerical
results for the correlated disorder, in the same fashion and for the same successive times as in Fig. \ref{fig2}.
Unlike Fig. \ref{fig2}, now all the data, including the low energy sites, fall on one smooth curve, given by
Eq. (\ref{stataver}) with $\beta=0.11$ and $\mu= -10$. Note that the spread in the site energies here is significantly
larger than in Fig. \ref{fig2}. This happens because in the convolution process $\epsilon_n$'s larger (in absolute
value) than the maximal "bare" value ($W=7$, in our case) emerge. This wide spread of $\epsilon_n$'s explains why
the chemical potential is so low. Indeed, if the nonlinearity were very small, then Eqs. (\ref{expand}) would imply
that $\mu$ must be considerably lower than the lowest site energy. Although in our case the nonlinearity is not
very small, nevertheless, the large spread in the site energies pushes the chemical potential down (provided, that 
the system does thermalize).

In Fig. \ref{fig3}c,d we plot the numerical results for the same disorder realization as in Fig. \ref{fig3}a,b 
but with a different initial preparation. These data display an interesting feature: the green rombs or the blue 
triangles arrange themselves nicely in two distinct curves, which merge into one for larger $\epsilon$'s. This 
suggests that we have here two subsystems of sites which thermalized separately, with two somewhat different 
$\beta$ and $\mu$. And then, in the course of time, there is a slow process of equilibration between the two 
subsystems. It is tempting to suggest that the points which form the upper or lower curves originate from sites 
belonging to one or the other half-chain. We have checked and confirmed this hypothesis. The asymmetry between
the two half-chains is due the fluctuations in $\epsilon$'s, as well as in the initial values of $I_n$'s. We 
have also studied some other disorder realizations from the aforementioned correlated ensemble. While the details 
of the time evolution depend on the realization and the initial preparation, we have observed in all examples 
an approach to equilibrium - in sharp contrast with the uncorrelated case.

{\it Conclusions --} We demonstrated that correlations in disorder strongly affects thermalization. We can offer 
the following intuitive explanation for this effect: In the absence of correlations (and for strong disorder) the 
mechanism of thermalization occurs via ``resonant triples'', i.e. clusters of three oscillators with close 
eigenfrequencies $\epsilon_n$ \cite{B10}. Correlated disorder enhances the probability of occurrence of such resonant 
triplets, thus, facilitating thermalization. We should stress again, however, that, since in our study the strong 
inequalities $\tau<<1, \rho<<1$ are not satisfied, we are not quite in the regime of a very weak coupling and 
nonlinearity, studied in Ref. \cite{B10}.

We are grateful to D. Basko for clarifications concerning Ref. \cite{B10} and to E.  Gurevish, D. Shepelyansky 
and A. Pikovsky for useful discussions. We are particularly indebted to D. Cohen for participation in the early 
stage of this work and for useful advise. This research was supported by grants from AFOSR No. FA 9550-10-1-0433, 
the DFG {\em Research Unit 760}, and the US-Israel Binational Science Foundation (BSF), Jerusalem, Israel.



\begin{thebibliography}{99}
\bibitem{note1}We note that in Ref. \cite{RCKG00}, the authors investigated thermalization in clean, i.e. 
disordered-free, lattices.
\end{thebibliography}
\end{document}